%% file: main.tex
\documentclass{article}

     \usepackage[preprint]{neurips}

\usepackage[utf8]{inputenc} 
\usepackage[T1]{fontenc}    
\usepackage{hyperref}       
\usepackage{url}            
\usepackage{booktabs}       
\usepackage{amsfonts}       
\usepackage{nicefrac}       
\usepackage{microtype}      
\usepackage{paralist} 

\usepackage[table, dvipsnames]{xcolor}

\title{Overcoming {\em Failures of Imagination} in AI Infused System Development and Deployment}

\author{
Margarita Boyarskaya \\
New York University \\
\And
Alexandra Olteanu  \\
Microsoft Research \\
\And
Kate Crawford \\
Microsoft Research \\
}

\begin{document}
\maketitle

\begin{abstract}
  NeurIPS 2020 requested that research paper submissions include impact statements on ``potential nefarious uses and the consequences of failure.''
  However, as researchers, practitioners and system designers, a key challenge to anticipating risks is overcoming what \cite{clarke1962hazards} called `failures of imagination.' 
  The growing research on bias, fairness, and transparency in computational systems aims to illuminate and mitigate harms, and could thus help inform reflections on possible negative impacts of particular pieces of technical work. 
  The prevalent notion of \textit{computational harms}---narrowly construed as either allocational or representational harms---does not fully capture the open, context dependent, and unobservable nature of harms across the wide range of AI infused systems. 
  The current literature focuses on a small range of examples of harms to motivate algorithmic fixes, overlooking the wider scope of probable harms and the way these harms might affect different {\em stakeholders}.
  The {\em system affordances} may also  
  exacerbate harms in unpredictable ways, as they determine stakeholders' control (including of non-users) over how they use and interact with a {\em system output}.
  To effectively assist in anticipating harmful uses, we argue that frameworks of harms must be context-aware and consider a wider range of potential {\em stakeholders}, {\em system affordances}, as well as viable proxies for {\em assessing harms} in the widest sense.
\end{abstract}

\section{Introduction}

\begin{quote}
    \em \scriptsize
    ``To see things as they really are, you must imagine them for what they might be.''~\cite{bell1995s}
    \vspace{-6pt}
\end{quote}

There is an increasing number of calls to put in place processes that require researchers, designers, and practitioners to reflect on, anticipate, and communicate possible failures and harmful effects from the technologies and the applications they develop or enable. 
Some of these calls have been realized through structured documentation and checklists aiming at codifying a range of responsible AI principles~\citep{arnold2019factsheets,mitchell2019model,gebru2018datasheets,sokol2020explainability,stoyanovich2019nutritional,madaio2020co,kiran2015beyond}, 
and more recently by the NeurIPS conference requirement for research paper submissions to include broader {\em social impact statements} on ``potential nefarious uses and the consequences of failure.''

However, making projections about risks, failures, and harms is by no means trivial. 
Foreseeing failures and harms that one has not observed before or that occur in new contexts is difficult even when it seems like they should have been predictable in hindsight. 
The eminent science-fiction author Arthur C. Clarke refers to these lapses as {\em failures of imagination}~\citep{clarke1962hazards}.
This phenomenon occurs even when the process of identifying risks is guided by extensive, well researched checklists~\citep{madaio2020co,kiran2015beyond,wong2020beyond}; since such checklists tend to be general and often fail to account for differences between technologies, applications, and stakeholders~\citep{kiran2015beyond}, while the norms that shaped their creation often remain hidden~\citep{lucivero2019making}. 
These might even contribute to shaping a `limited’ imagination of 
what sort of harms we consider, 
when we consider them, 
how we operationalize and measure them, and 
what trade-offs we make when attempting to mitigate them.
But are there any tools that can help AI practitioners do better? 
What makes AI systems different from other technologies?
What can we learn from the literature on responsible innovation~\citep{stilgoe2013developing, grunwald2011responsible}, technology assessment~\citep{kiran2015beyond}, critical race theory~\citep{bell1995s, benjamin2019race} or feminist theory~\citep{costanza2018design, d2020data}?

More often than not the existing scholarship on harmful uses of technology \citep{washington2020whose,taddeo2016debate,bijker1997bicycles} does not capture the complex nature of many AI systems that aim to meet user needs for e.g., information, social connection, or entertainment; 
nor does it address the variety of stakeholders and the many ways they interact with these systems.
Take the example of information retrieval systems that curate, rank, recommend, extract or represent information, among other uses.
Retrieval algorithms govern the interaction between systems and humans, determining {\em whose} information to show {\em to} individual users, and what information to display {\em about} individuals (possibly distinct from the user). 
These interactions might lead to a wide spectrum of negative overtones or impacts
for human stakeholders, ranging from denigration and emotional distress to physical harm and loss of opportunities \citep{blodgett2020language,barocas2017problem,kay2015unequal,abbasi2019fairness,baker2013white,otterbacher2017competent}.

To understand the effectiveness of the `broader impact' statement practice introduced by NeurIPS 2020, it is thus useful to examine the critique within the literature on `responsible innovation' where the very possibility of such innovation has been challenged~\citep{blok2015emerging,grinbaum2013responsible}.
One obstacle is the absence of a \textit{consensus} on the scope of so-called `wicked problems' and the goal of the innovation process among stakeholders.\footnote{\textit{Wicked problems} are characterized by having no definitive formulation, while their solutions are not true-or-false, but \textit{better} or \textit{worse} to some.  
Mitigating biases or unfairness in AI systems is an example of a `wicked problem,' in contrast to more tangible challenges like gaming algorithms for personal gain.}  
Another challenge relevant to anticipatory statements on `broader impacts' was captured as a principle by \cite{collingridge1982social}: ``The social consequences of a technology cannot be predicted early in the life of the technology. [...] This is the dilemma of control. When change is easy, the need for it cannot be foreseen; when the need for change is apparent, change has become expensive, difficult, and time consuming.''

With these difficulties in mind, we underscore the need for a wider discussion, interdisciplinary teams, and tools to assist in anticipating adverse consequences.
%
%
We conjecture that the failure to anticipate harms is often the result of researchers 
and practitioners' own failures of imagination and ask what, if anything, can be done to mitigate this. 
Drawing from a collection of pre-prints of NeurIPS'20 papers, we discuss how more context-aware frameworks of harms can contribute to a wider imaginative scope. 
Yet, to serve as an aide for \textit{exploration} and \textit{discovery}, we should consider a wide range of 
1)~stakeholders, system affordances, uses, and outputs, as well as 
2)~the characteristics and types of harm, and viable proxies for assessing them---going beyond the familiar `checklist' approach.

\vspace{-12pt}
\section{Overcoming Failures of Imagination}

\vspace{-8pt}
\begin{quote}
    \em \scriptsize
    ``Radical assessment can encompass illustration, anecdote, allegory, and imagination, as well as analysis of applicable doctrine and authorities.''  
    -- \cite{bell1995s}
    \vspace{-8pt}
\end{quote}

The discussions about the safety of technology outcomes, including for existing or possible research applications, tend to be dominated by how to forecast or identify failures.
\cite{clarke1962hazards} 
describes two ways by which forecasting can fail: failure of nerve and failure of imagination. 
Failure of nerve prevents us from imagining entirely new possibilities (discoveries, technologies). 
The current popularity of AI has rendered this first warning of Clark less poignant: a series of notable advances, the excited discourse, and the widespread adoption of AI have accustomed both researchers and users of technology to thinking about virtually every problem as amenable to some sort of AI solution. 

The second `hazard of forecasting'---the failure of imagination---refers to insufficient, impoverished visions of the future that do not adequately capture the complexity of the upcoming reality. 
While `failures of nerve' might arguably be of less concern, fast and short research-to-application pipelines and extensive scope of AI impact have made the failures of imagination more material.
At the same time, there is a movement to bring harm minimization closer to the stage of technology \textit{design} rather than post-\textit{deployment} \citep{selbst2019fairness}, 
with the duty of imagining the consequences of technology being placed on the researchers and developers themselves. 


\input{table}

\textbf{Understanding current practices.}
To understand how the `broader impacts' call was construed, we examined pre-prints of NeurIPS'20 accepted papers,\footnote{We surveyed 35 papers that were available on arXiv.org before the NBIAIR workshop submission deadline. Some of the reviewed impact statements might have been updated for camera ready versions of these papers.}   
Table~\ref{table:statement_themes_examples} overviewing emerging themes.

Broadly, we observe that even \textit{acknowledging  uncertainty} in imagining adverse consequences of technology can itself be a pitfall. 
%
%
Authors commonly assess impact only through the narrow lens of specific technical contributions.
For example, authors might position work on preventing adversarial attacks as innately positive and fail to deliberate about limitations and unintended harms. 
Another concerning theme is overlooking the harms and interests of certain (often disadvantaged) \textit{stakeholders}. 
%
This is often the result of
only considering `benefits' to those developing and deploying the technology and `harms' as synonymous with harm to a company, or some type of mass disaster.
The {\em severity} and importance of harms to individuals is also often ignored.


\textbf{Context-aware frameworks of harm.} 
Avoiding a broader range of harms is difficult as they might require different optimization and design goals, with some types of harm also being harder to operationalize than others. 
A framework for systematically reasoning about harms could, however, assist practitioners in probing them across applications and usage scenarios. 
Drawing from the literature on risk perception, computational fairness, and psychology of decision making, we briefly highlight contextual idiosyncrasies with respect to stakeholders, usage affordances, and system response that may affect the (perceived) {\em salience}, {\em severity}, and {\em attribution} of harms. 


\textit{The stakeholders.}
Those at risk of being affected are central to understanding potential harms.
Their traits and circumstances can help delineate various types of harms.
Affected stakeholders might be neither users of a system under examination (e.g., recruiters on LinkedIn) nor those developing it, but also subjects being rated or ranked by a system 
(e.g., an academic ranked on Rate My Professors)
or content producers 
(e.g., musicians on Spotify).
Stakeholders {\em vulnerability} (propensity to be affected) and {\em agency} (degree of control over a system behavior and their interaction with it) may mediate the salience and severity of harm. 
In addition, demographic cues can also activate 
stereotypical beliefs and affect the imagination of what harms are possible.


\textit{System affordances.}
The idiosyncrasies of a system (e.g., search engine versus specialized prediction software) and of particular usage scenarios can also provide cues about the range of possible harms. 
The mechanisms through which stakeholders interact with a system may determine what harms are more likely and who might be affected.
For instance, while restricted access to some critical service for some stakeholders \citep{abbasi2019fairness} affects them directly,
the exposure or representation {\em of} stakeholders 
\citep{biega2018equity,singh2018fairness} may also affect those creating or consuming the content.

\textit{System use and response. }
It is critical to consider 
under what circumstances harm might occur or could be actualized, such as
{\em what} may be affected (e.g., well-being, opportunities, dignity),
{\em how} it may happen (e.g., immediately, frequently, by altering beliefs or by taking an action), and 
{\em why} it may happen (e.g., due to a process or an outcome). 
Problematic system behavior and outputs may lead to harms if there is any threat 
of unjust resource {\em allocation} or of opportunity loss;  
of unfair {\em representation} of someone in the system output, such as reinforcing stereotypes about a group \citep{abbasi2019fairness,crawford2017trouble}; 
or to stakeholders' {\em agency} or autonomy (e.g., nudged towards certain actions or beliefs);  and their {\em physical} or {\em emotional well-being} (e.g., trauma, anxiety).

\vspace{-4pt}
\section{Towards Responsible AI Innovation}
\vspace{-8pt}

While the introduction of impact statements to NeurIPS 2020 is an important step towards making evaluations of harm a more consistent and effective research practice, some emerging patterns merit further examination.
By outlining challenges to deliberating about harms as part of broader impact statements and describing elements that determine \emph{who} is affected and \emph{how} they are affected, ~we ask what can be done to address possible failures of imagination.
Rather than suggesting a definitive approach, we hope this will draw attention to the need for context-aware frameworks that can help anticipate harmful outcomes of AI systems.

\cite{stilgoe2013developing} highlight four dimensions of responsible innovation: anticipation, reflexivity, inclusion, and responsiveness.
We suggest probing the limits of harm \textit{anticipation} by thinking systematically about sociotechnical affordances, use context, and the conflicting interests of various stakeholders.
%
\textit{Reflexivity} means questioning one's own assumptions about e.g., the neutrality or the benefits of technology adoption (a persistent assumption among NeurIPS authors).
Beyond hypothesizing about stakeholders, true \textit{inclusion} cannot be achieved if the intended beneficiaries are not part of the conversation. Soliciting input from those that may be affected by the applications of research work is thus critical.
A natural follow-up to admitting the limits of one's knowledge is to also engage with domain experts (e.g. social scientists, social workers) over time, including both during the \textit{envisioning} (model design) and the \textit{response} (mitigation) stages.
These dimensions 
particularly challenge two concerning trends present in the broader impact statements (Table~\ref{table:statement_themes_examples}):


\textbf{Challenging the technological benevolence frame.}
Construing ethics as strictly the responsibility of practitioners and users (and not that of researchers) is a persistent and notable theme across the NeurIPS'20 impact statements.
Changing research norms can however impact practice in the industry and beyond. 
Some of the `broader impact' statements we examined suggest that researchers often presume the \textit{neutrality of general-purpose algorithms}. 
However, even generic, base models might be developed with particular applications and stakeholders in mind, which may result in performance disparities. 
One example are generic sentiment classifiers initially developed to assess product reviews but later also used to predict e.g., suicidal intent. 
%
Even with some mitigation in place, there are also implications of whether the process by which potential harms are considered is \textit{prescriptive} (i.e., allow lists) or \textit{restrictive} (i.e., block lists). 
While researchers adopting a prescriptive approach may recognize their inability to predict harmful scenarios by limiting the use of their work, the applications they deem safe may not serve the needs of all stakeholders.  
Reflecting on the implicit assumptions of each approach is a necessary step towards probing the limits of one's concept of harm.

\textbf{Reflecting on who is vulnerable.}
Some NeurIPS'20 authors also appear to overlook the more vulnerable stakeholders in their deliberations of broader impacts.  
The notions of vulnerability and harm are defined in relation to specific stakeholders and are inseparable from the sociotechnical context \citep{bijker1997bicycles}. 
This is why a better understanding of human perceptions of harm and what impacts these perceptions is necessary to identify and categorize problematic practices, as well as to obtain insights into users' choices around how and when to use certain systems~\citep{skirpan2018s}.
It can also assist in determining {\em whether} and {\em when} such harms are foreseeable and {\em by whom}. 

\noindent {\bf Acknowledgements:} We thank Solon Barocas for insightful discussions and feedback.

\footnotesize
\bibliographystyle{ACM-Reference-Format}
\bibliography{references}

\end{document}

%% file: table.tex
\begin{table}[t!]
\scriptsize
    \centering
    \def\arraystretch{0.9}
    \begin{tabular}{@{}p{6.6cm}|p{7cm}@{}}
    {\textbf{Themes}} &  
    {\textbf{Example quotes}} \\\hline
    \rowcolor{lightgray}
    \multicolumn{2}{@{}l}{\textbf{Examples of concerning trends}} \\
    \textit{Neglecting  stakeholders} (typically minority groups), by assuming
    
	- `benefits' to mean benefits to companies, governments, etc. 
	
	- `harms' to mean impediments to technology deployment 
	
	- `bad intentions' to be of users and not of technology developers
	
	- `harm' only in relation to war, government, or `mass disaster'
	
	- or using terms like `reliable', `secure' without specifying \textit{for who}
	& 
	``our work can bring both beneficial and harmful impacts and it really depends on the motivation of the users'' \protect{\citep{hu2020one}} and  
	 ``[the work is] academic in nature, and does not pose
foreseeable risks regarding defense, security, and other sensitive fields.'' \protect{\citep{aksan2020cose,wang2020pie,jiang2020shapeflow}} 
	\\\hline
    \textit{Outsourcing the ethical responsibility} 
	to others or other stages of technology deployment by ignoring theoretical or technical affordances for misuse and instead referencing biased inputs, engineering mistakes or malicious uses 
     &  ``there exist risks that some engineers [can] deliberately use the algorithm [in a way that would] harm the performance of the designed system'' \protect{\citep{hu2020one}} \\\hline
    \textit{Confusing technical advances with positive impact}, by 
    
    - assuming adoption of technical solutions to constitute a benefit 
    
    - failing to question assumptions behind performance metrics 
    
    - treating impact statement as a `sales pitch' 
	& 
	``Further extensions include applying
[the method] in robotics. Machine learning for robotics is increasingly growing as a field and has potential of revolutionizing technology in the unprecedented way.'' \protect{\citep{choromanski2020ode}} 
	\\\hline
	%
    \textit{Suggesting the research topic bounds the scope of inquiry}
         (e.g., fairness papers failing to acknowledge limitations or possible unintended negative effects, 
         theory papers suggesting they are exempt from reflections on impact) 
         & ``[this work] is theoretical and conceptual in nature and so is its likely current broader impact'' \protect{\citep{vialard2020shooting}}; ``[the] study is crucial as it indicates the vulnerability of [DNN] classifiers to adversarial attacks'' \protect{\citep{dolatabadi2020advflow}}
    \\\hline
    \textit{Emphasizing the net impact of the paper}, (e.g. defending, de-emphasizing, or `balancing' harms with unrelated benefits) 
    & ``the positive impact of foundational research on public datasets, such as is presented in this paper, far outweighs [risks] lying further downstream.'' \protect{\citep{asano2020labelling}}
    \\\hline
	\textit{Overconfidence and not acknowledging epistemic uncertainty} (e.g. ascertaining no harm).
	&
	``if the method fails in some extreme circumstances, it will confuse researchers or engineers [but] it will not bring about any negative ethical or societal consequences.'' 
	\protect{\citep{wang2020transferable}} \\
	
	\rowcolor{lightgray}
	\multicolumn{2}{@{}l}{\textbf{Examples of encouraging trends}} \\
	 \textit{Considering a variety of stakeholders}, including by situating harms on micro and macro levels (environmental, economic, individual)
	& ``We summarize the potential impact [to] the research community[, to] the downstream engineers[, to] the society'' \protect{\citep{hu2020one}} 
	\\\hline
	\textit{Admitting uncertainty} by 
	
	- stating epistemic limitations to envisioning possible impacts
	
	- promoting involvement of and collaboration with domain experts
	&
	``[this] work is primarily theoretical, making its broader impact difficult to ascertain'' \protect{\citep{pfau2020disentangling}}; 
	``[performing] meaningful probabilistic inferences requires [domain] expertise regarding modeling priors and potential biases of [sampling approximations]'' \protect{\citep{pleiss2020fast}}
	\\\hline
	
	%
	 \textit{Deliberating even about the risks of mitigation strategies} by 
	 
	 -  examining known fairness or accountability issues (e.g., whether better interpretability can lead to harm by advancing stereotypes)
	 
	 - examining weaknesses of technological improvements
	& ``failure [to calibrate the model can] lead to poor risk management in high risk applications''  and ``black-box optimization [related risks include] over-reliance on fully automated methods and computationally expensive searches [for] marginal improvements.'' \protect{\citep{pleiss2020fast}}
	\\\hline
	\textit{Giving examples} (tasks, failure scenarios, situations of harm)
    & 
    ``[could] adversely impact people employed via crowd-sourcing sites [and] manually review and annotate data'' \protect{\citep{sharma2020self}}
	\\\hline
    \end{tabular}
    \caption{Example themes present in broader impact statements from NeurIPS'20 papers.}
    \label{table:statement_themes_examples}
    \vspace{-22pt}
\end{table}